\documentclass[a4paper,20pt]{article}
    \usepackage[T1]{fontenc}
    \usepackage{graphicx}
    \usepackage{epsfig}
    \usepackage{amsmath}
    \usepackage{amsfonts}
    \renewcommand{\abstract}{}
\begin{document}
\makeatletter
\renewcommand{\@oddhead}{\textit{YSC'14 Proceedings of Contributed Papers} \hfil \textit{O.V. Zakhozhay, A.P. Vid'machenko, V.A. Zakhozhay}}
\renewcommand{\@evenfoot}{\hfil \thepage \hfil}
\renewcommand{\@oddfoot}{\hfil \thepage \hfil}
\fontsize{11}{11} \selectfont

\title{Simulation of the Continuous Spectrum of Substars with Protoplanetary Discs}
\author{\textsl{O.V. Zakhozhay$^{1}$, A.P. Vid'machenko$^{2}$, V.A. Zakhozhay$^{1}$}}
\date{}
\maketitle
\begin{center} {\small $^{1}$V.N. Karazin Kharkiv National University \\
$^{2}$Main Astronomical Observatory NAS of Ukraine \\
zkholga@mail.ru, qwe-qwe@ukr.net}
\end{center}

\begin{abstract}
The continuous spectra of the substars with surrounding
protoplanetary disks were calculated. The results reveal that
protoplanetary disc average temperature decreases to 3 K during the
period of 5 Myr for substars with masses 0.01$M_{_{\bigodot}}$ and
during the period of 160 Myr for substars with masses
0.08$M_{_{\bigodot}}$. Estimations of protoplanetary discs flux
maximum depending on the substar mass at the age of 1 Myr are: 4.6
kJy (for 0.01$M_{_{\bigodot}}$) and 3.4 MJy (for
0.08$M_{_{\bigodot}}$). Maximum of protoplanetary disc radiation
before it reaches the temperature of the cosmic microwave background
changes within the ranges: from 0.07 mm to 0.58 mm (for substar mass
0.01$M_{_{\bigodot}}$) and  from 0.02 mm to 0.29 mm (for substar
mass 0.08$M_{_{\bigodot}}$).
\end{abstract}

\section*{Introduction}
\indent \indent Substars (Brown Dwarfs) are celestial bodies, which
are formed by self-gravity and evolve due to internal energy
resources. Such young bodies are found in the young open clusters in
the galactic field and as invisible components in the binary
systems. Recently the excess of Infrared (IR) radiation in the young
substars was found \cite{1liu}, which were interpreted as a
protoplanetary disc existence.

\section*{Model}

The aim of this research is to calculate the radial flux from
substars with protoplanetary discs, which is expected in the first
milliard year of their existence.

To solve this problem the following model was considered:
\begin{itemize}
  \item substars masses are within the range 0.01$M_{_{\bigodot}}$..0.08$M_{_{\bigodot}}$;
  \item protoplanetary discs lie in the plane of the sky;
  \item radii of protoplanetary discs are equal to 100 a.u.;
  \item substars and protoplanetary discs radiate as black body;
  \item substars and protoplanetary discs radiate like black body;
  \item protoplanetary discs effective temperature equals to average
temperature of the protoplanetary disc;
  \item average temperature reduction is the consequence of substar
luminosity changing;
  \item distance from the Sun to a substar equals to 10 pc.
\end{itemize}

For our calculations we used the results from the paper
\cite{2pisarenko} for substars cooling simulation. For estimation of
the average temperature of protoplanetary discs the results from
\cite{3saphronov} were used. They describe the dependence between
the temperature gradients and the distance to a central source which
has current radius and effective temperature and also depends on
disc thickness. It was assumed that the mechanism of protoplanetary
disc heating by stars and substars are the same.

\section*{Results and Discussion}

According to Saphronov \cite{3saphronov} and Larson \cite{4cameron}
calculations, changing of the protoplanetary cloud temperature (with
different radius dependences of its thickness) with distance to
central body has similar changing and has a correlation with disc
thickness. In the areas which has the same distance from central
source of heating a higher temperature corresponds to thicker discs
(Fig. 1). It gives possibility to use Saphronov \cite{3saphronov}
equation for temperature gradient in the cloud:
\begin{equation}
\label{eq1} T_R \propto \left( {\frac{R_{ss} }{R}}
\right)^{3/4}T_{eff} ,
\end{equation}
Where $T_{R }$ is the disc temperature on the distance R from the
center of the substar which has radius $R_{ss}$.

\begin{figure}[t]\centering
\begin{minipage}[t]{.75\linewidth}
\centering \epsfig{figure=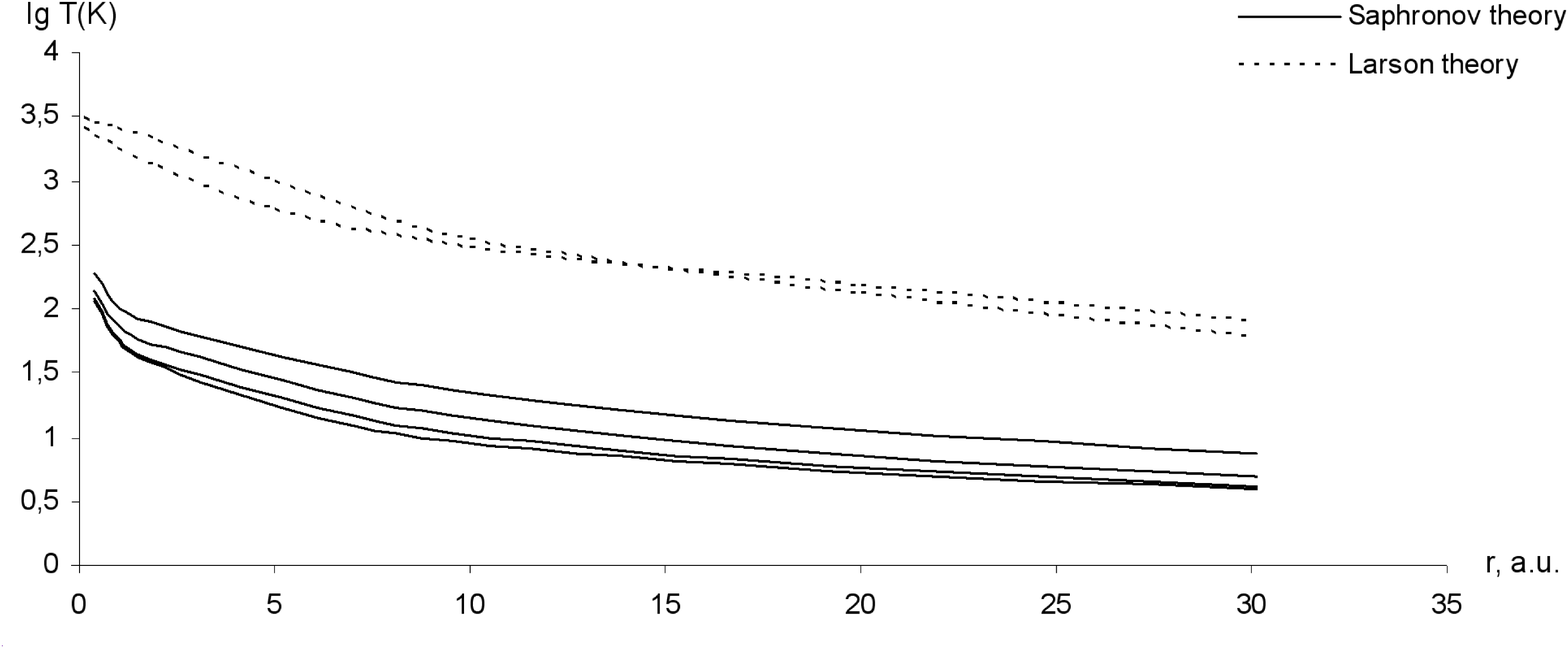,width=\linewidth}
\caption{Temperature distribution from the central plane of the
protoplanetary disk models according to calculations of Larson
\cite{4cameron} and Saphronov's model \cite{3saphronov}.}
\end{minipage}
\end{figure}

Integration of the equation (\ref{eq1}) from substar radius to disc
radius gives the disc average temperature
\begin{equation}
\label{eq2} \bar {T}_d =\frac{1}{R_d }\int\limits_{R_{ss} }^R
{T_R dR} ,
\end{equation}
Taking into account that substar radius is considerably less than
protoplanetary disc radius $R_{ss}\ll R_{d}$ and includes the
proportional coefficient $C$, the equation for average disc
temperature is the following
\begin{equation}
\label{eq3} \bar {T}_d =CT_{eff\_ss} \left( {{R_{ss} }
\mathord{\left/ {\vphantom {{R_{ss} } {R_d }}} \right.
\kern-\nulldelimiterspace} {R_d }} \right)^{3/4},
\end{equation}
For estimation of $C$ we used the data for Vega luminosity
(temperature and star radius) from the catalog of the nearest stars
\cite{6zakhozhaj}. We also assumed the parameters of protoplanetary
disc - average temperature 85~K and size 100 a.u.. Thus we obtained
$Ñ\approx 10$.

Using temperature (3) as an effective disc temperature the
dependence of the protoplanetary disc flux and central source on
radius and effective temperature of substar were calculated. The
disc whose internal radius is equal to the radius of the substar was
chosen as the radiation zone of the protoplanetary disc. We assume
that it radiates as black body. In this model the disc thickness is
characterized by the coefficient $C$ calculated above.

Figures 2 and 3 show the sum of fluxes from the substars with masses
$0.08M_{_{\bigodot}}$ and $0.01M_{_{\bigodot}}$ respectively and
protoplanetary disc with size of 100 a.u. and distances 10 pc. Their
left parts represent the continuous spectrum radiation from the
substar and the right part - mainly from protoplanetary disc. As the
temperature and the radius of the substars change simultaneously
during evolution the maximum of the substar radiation moves to the
right. This process causes disc cooling: the radiation maximum moves
toward the longer waves.

\begin{figure}[!ht]
\centering
\begin{minipage}[t]{.75\linewidth}
\centering \epsfig{figure=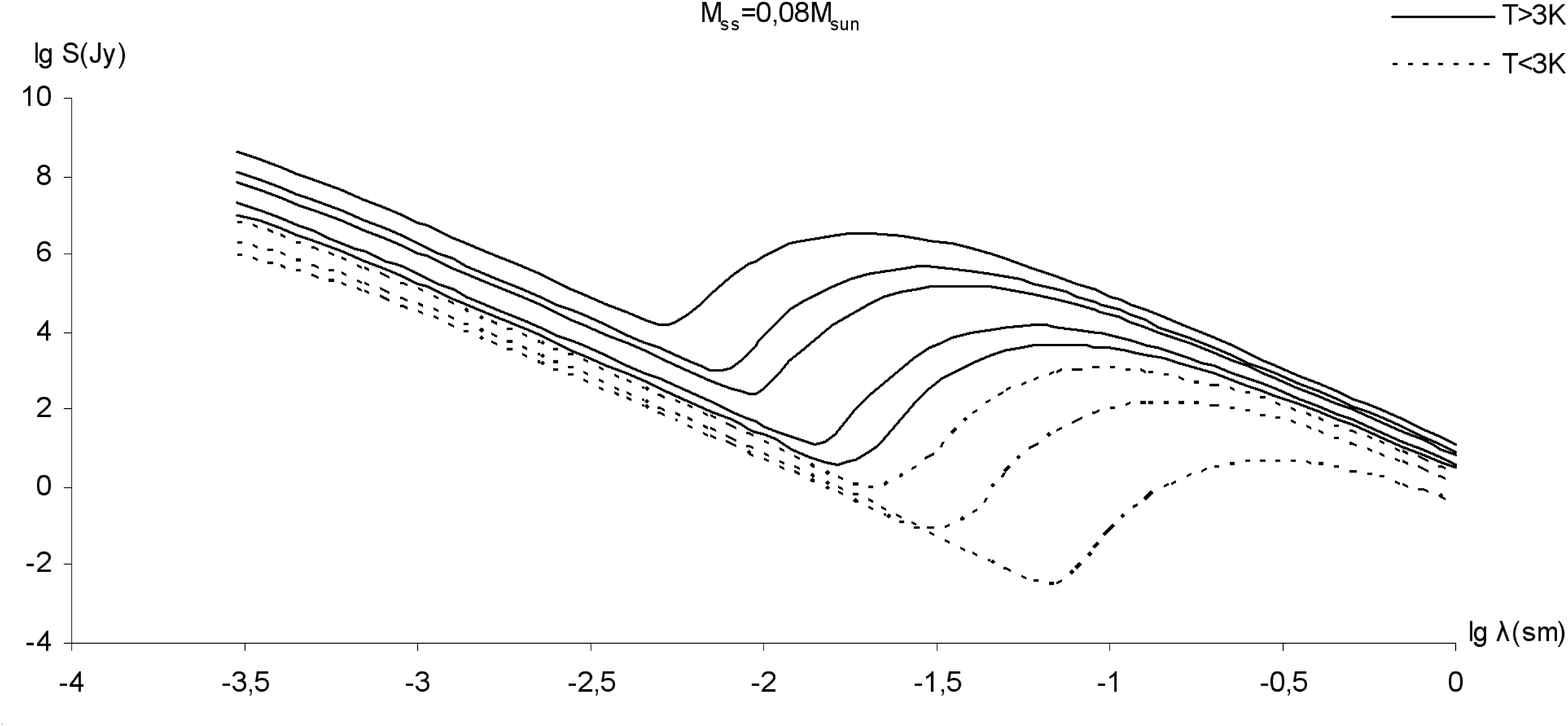,width=\linewidth}
\caption{Distribution of the total flux density from substar with
mass equals to $0.08M_{\bigodot}$ and accepted model of the
protoplanetary disk for different age t (years). The top curve
corresponds to $\log(t)=6$, the bottom one - to $\log(t)=9$.}
\end{minipage}
\begin{minipage}[t]{.75\linewidth}
\centering \epsfig{figure=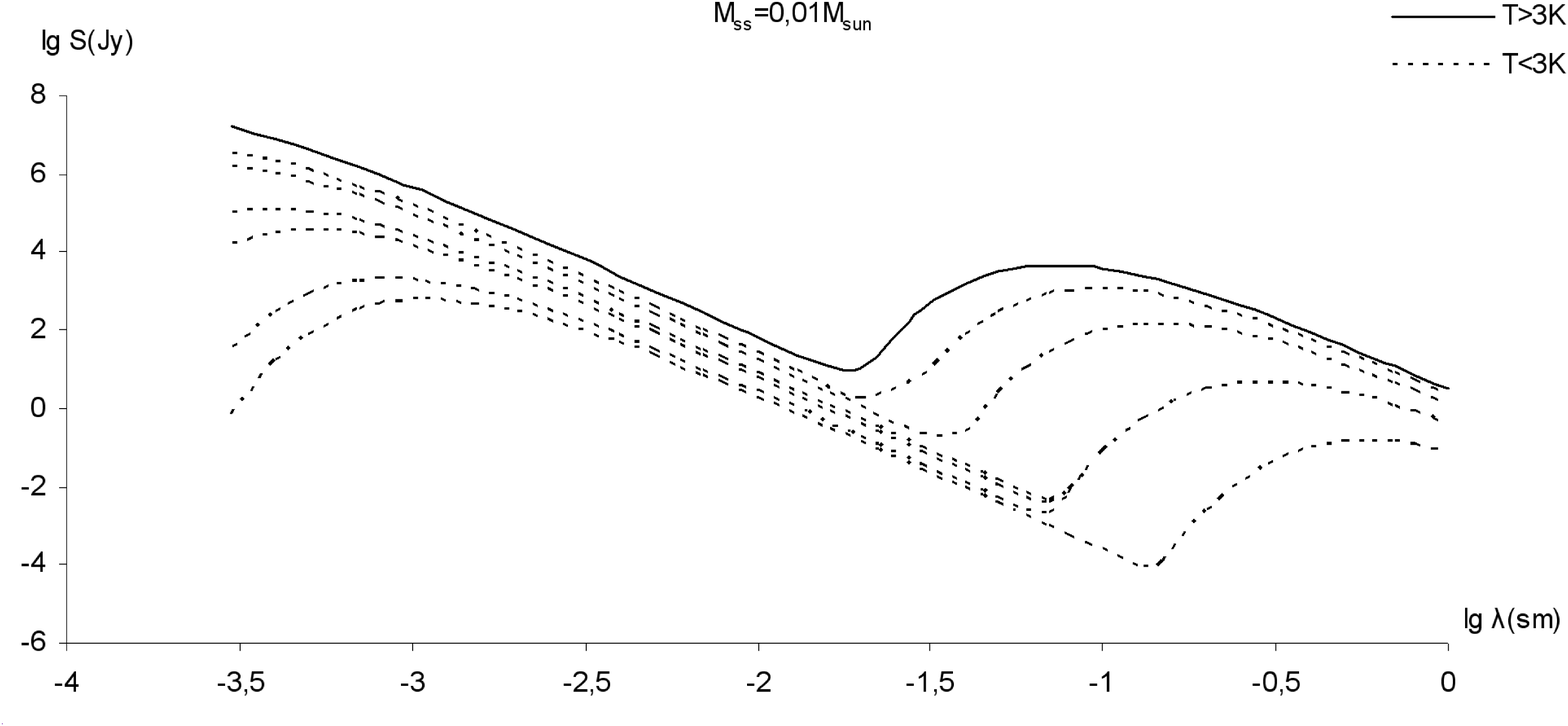,width=\linewidth}
\caption{ Distribution of the total flux density from substar with
mass equals to $0.01M_{\bigodot}$ and accepted model of the
protoplanetary disk for different age t (years). The top curve
corresponds to $\log t=6$, the bottom one - to $\log t =9$.}
\end{minipage}
\end{figure}

Theoretically there is a possibility to discover disc until its
temperature is higher than microwave background temperature (2.7K).
According to calculations average temperature of the protoplanetary
disc reaches 3K during 5 million years for substars with masses
$0.01M_{_{\bigodot}}$ and during 160 million years for substars with
masses $0.08M_{_{\bigodot}}$. Thus radiation stage for
protoplanetary discs when their fluxes are higher than microwave
background is quite short.

The calculations also allow to see that protoplanetary discs flux
maxima shift to submillimeter and millimeter wavelength range. In
this time substars must be seen in IR. It was found that the flux
from protoplanetary discs has maximum just in millimeter range and
equals to $\sim3.4$MJy for spatial models.

Fig. 4 represents the flux and the wavelength changing with mass
which corresponds to maximum radiation from the protoplanetary
discs. The left part of the plot corresponds to the flux from the
star (from $0.08M_{_{\bigodot}}$ (left) to $0.01M_{_{\bigodot}}$
(right). The right part of the plot corresponds to the flux from the
protoplanetary disk (all the curves coincide).
\begin{figure}[!ht]\centering
\begin{minipage}[t]{.75\linewidth}
\centering \epsfig{figure=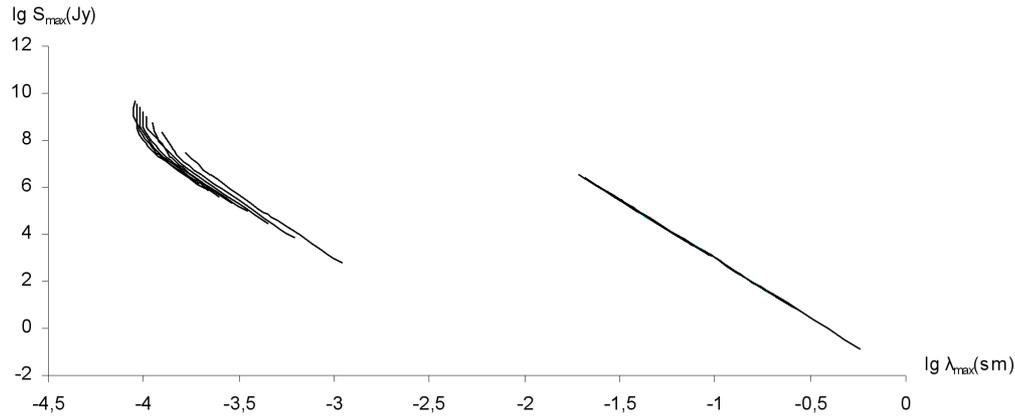,width=\linewidth}
\caption{The wavelength dependence of the flux changing which
corresponds to maximum radiation from the protoplanetary discs with
time}
\end{minipage}
\end{figure}

It is seen that fluxes in substars maximum radiation and their
protoplanetary discs changes according to different laws. Within
this time ranges substars fluxes decrease according to parabolic
dependence, the fluxes from discs decrease linearly according to
equation
\begin{equation}
\label{eq4} \lg S=-5\lg \lambda -2.033
\end{equation}
In the star formation regions one could detect the excess of the
flux from stars and substars which is due to presence of protostars.
As the Sun is situated in the place of intensive stars formation the
comparison of fluxes obtained by IRAS (Infrared Astronomical
Satellite) and presented in the catalogue of the nearest IR-sources
\cite{7zakhozhaj} and in catalogue of the nearest stars
\cite{6zakhozhaj} which consist of late M-dwarf, i.e. substars, was
done. As a result 6 sources of IR -- radiation from Zkh 7Ñ, Zkh
162Â, Zkh 235Ñ, Zkh 249-250, Zkh 350, Zkh 353 systems were
discovered. These systems may be examined as candidates of substars
with protoplanetary discs. The sensitivity of modern
radiation-measuring instruments in millimeter wave range is enough
for their identification.

\section*{Conclusions}
\indent \indent As it follows from the results the protoplanetary
discs of the substars are available for detection when their age is
less then first ten million years. The mentioned nearest late
M-dwarfs with IR excess and the estimations of radiation fluxes of
the protoplanetary discs in millimeter wave range allow to put a
question about real detection of the discussed objects in the
circumsolar region.

\end{document}